\documentclass[aps,prb,amsfonts,showpacs,floats,floatfix,twocolumn,nofootinbib,byrevtex]{revtex4}
\usepackage{epsfig}

\begin{document}

\title{Shape dynamics during deposit of simple metal clusters on rare gas matrices}
\author{P.~M.~Dinh$^1$, F.~Fehrer$^2$,  G.~Bousquet$^1$,
  P.-G.~Reinhard$^2$, and  E.~Suraud$^1$}
\affiliation{
$^1$Laboratoire de Physique Th\'eorique, UMR5152, Universit{\'e} de Toulouse, 
  118 Route de Narbonne, F-31062 Toulouse C\'edex, France \\
$^2$Institut f\"ur Theoretische Physik, Universit{\"a}t Erlangen,
   Staudtstrasse 7, D-91058 Erlangen, Germany 
}
\begin{abstract}
Using a combined quantum mechanical/classical method, we
study the collisions of small Na clusters on large Ar clusters as a
model for cluster deposit. We work out basic mechanisms by systematic
variation of collision energy, system sizes, and orientations.  The
soft Ar material is found to serve as an extremely efficient shock
absorber. The collisional energy is quickly transfered at first impact
and the Na clusters are always captured by the Ar surface.
The distribution of the collision energy into the Ar system
proceeds very fast with velocity of sound. The relaxation of shapes
goes at a slower pace using times of several ps.
It produces a substantial rearrangement of the Ar system while the
Na cluster remains rather robust.
\end{abstract}
\maketitle
\section{Introduction}
\label{s:intro}

Clusters on surfaces have been much investigated during the past
decade as can be seen, e.g., from the sequence of recent ISSPIC
proceedings \cite{ISSPIC9,ISSPIC10,ISSPIC11,ISSPIC12}. The topic
remains of great interest, especially in relation to the synthesis of
nano-structured surfaces. A possible way is here the direct deposition
of size selected clusters on a substrate \cite{Bin01,Har00}. When
deposited on a surface, a cluster undergoes a significant modification
of its electronic structure and ionic geometry, because of the
interface energy, the electronic band structure of the substrate, and
the surface corrugation.  These questions have been widely
investigated from the structural point of view both at the
experimental \cite {Exp1,Exp2,Exp3} and theoretical
\cite{BL,CL,HBL,MH,SurfPot1,SurfPot2,2harm,IJMS} sides. The
theoretical description of deposition dynamics is nevertheless a very demanding
task.  A fully microscopic approach treating all atoms and electrons
in detail is extremely involved and applicable only to very small
samples if at all. Most theoretical descriptions thus employ simple
molecular dynamics approaches with effective atom-atom forces.  There
are, however, situations where a detailed description of electronic
degrees of freedom is desirable, for example when metal clusters are
involved where electronic shell effects are known to play a role in
forming the structure.

A step forward is to combine a fully microscopical treatment of the
cluster with a much simplified description of the surface. This is a
valid and efficient compromise for inert substrates as, e.g., for the
deposit of a metal cluster on an insulator surface. Such an approach
was explored, e.g., in the case of Na clusters on NaCl in
\cite{IJMS,Ipa04} for which Density Functional Theory was used for the
Na electrons coupled to the surface via an effective interface
potential, itself tuned to ab-initio calculations \cite{MH}.  This
approach, however, ignores any excitation and/or rearrangement of the
surface itself, which supposes extremely inert materials. The next
step is obviously to restore a minimum of surface degrees of freedom,
still having electronically inert substrate in mind. Such a model was
recently developed for the case of Na clusters embedded in rare gases
\cite{Ger04b,Feh06a,Feh05c} where the cluster is again treated in full
detail and the rare-gas atoms through classical dynamics of position
and dipole polarization. Such a "hierarchical" approach is justified
for electronically inert substrates. It has much in common with the
coupled quantum-mechanical with molecular-mechanical method (QM/MM)
often used in bio-chemistry \cite{Fie90a,Gao96a,Gre96a}.
The now explicit, although simple, handling of substrate atoms
still restricts calculations to finite systems. The model developed in
\cite{Feh05c,Feh06a,Feh06b}, nevertheless, allows to explore a
sufficiently large range of sizes to see the appearance of generic
behaviors on the way towards the bulk.
%

The goal of this paper is to study the dynamics of deposition of a
small Na cluster on a finite model of an 
Ar surface. To that end, we consider Na$_6$,
Na$_7$, or Na$_8$ clusters as projectile on Ar$_N$ clusters of various
sizes ($N$ between 7 and 87) as target.
We analyze the kinetics of deposition in terms of the energy
transfer between cluster and substrate, a point for which the explicit
dynamical surface degrees of freedom are crucial.
We shall also analyze the way in which the metal cluster adapts to the
substrate and how the substrate gives way.  This is done by tracking
the evolution of the shapes of both partners during the deposition
process and by varying Ar cluster size as well as initial kinetic
energy. For this purpose we shall compare several
small Na clusters with different shapes in the initial state.  
It should finally be noted that the considered setup (Na cluster on
Ar$_N$) has two aspects. First, it is an example of cluster-cluster
collisions involving partners of very different size and nature, as
such a topic of its own interest. Second, we have in mind an
exploratory study of deposition on a Ar surface for which we consider
large, but finite systems, a method which has been used also in
experimental studies, see e.g. \cite{Buc94a}. The widely used
alternative is to simulate the infinite surface in terms of a small,
but periodically copied, simulation box, see e.g. \cite{Xir02aR}. This
better allows to include long range effects but limits the structural
rearrangement of the surface. Thus modeling the surface by finite
clusters is probably more appropriate for our intention to analyze the
shape dynamics.

The paper is organized as follows. 
Section \ref{s:model} provides a short presentation of the model and
of the systems considered in the test cases.
In section \ref{s:Na6}, we analyze the deposition dynamics
in terms of trends and energies.
In section \ref{s:wet}, we discuss the shape dynamics in more detail.

\section{Model}
\label{s:model}

As stated above we use a hierarchical approach in which the metal
cluster is treated in full microscopic detail while the "surface" is
described at a classical level granting each atom mobility as a whole
and a dynamical dipole response.  This is justified by the clear
hierarchy of electronic binding in metals vs. rare gases.  The many
different ingredients make a complete description of the model rather
bulky. We give here a short account of the approach and refer to
\cite{Ger04b,Feh05c} for a detailed layout.

The Na cluster is treated using quantum-mechanical single-particle wavefunctions
$\{\varphi_n({\bf r},t),n=1,\ldots,N_{\rm el}\}$,
for
the valence electrons coupled non-adiabatically to classical
molecular dynamics (MD) for the positions of the Na ions
$\{{\bf R}_I,I=1,\ldots,N_{\rm Na}\}$.
The electronic wavefunctions are propagated by time-dependent
local-density approximation (TDLDA).
The electron-ion interaction in the cluster is described by soft,
local pseudo-potentials. This TDLDA-MD has been validated for linear
and non linear dynamics of free metal clusters \cite{Rei03,Cal00}.

Two classical degrees of freedom are associated with each Ar atom:
center-of-mass
$\{{\bf{R}}_a,a=1,\ldots,N_{\rm Ar}\}$, 
and electrical dipole moment 
$\{{\bf d}_a,a=1,\ldots,N_{\rm Ar}\}$.
With the atomic dipoles, we explicitely treat the dynamical
polarizability of the atoms through polarization potentials
\cite{Dic58}. Smooth, Gaussian charge distributions are used for Ar
ionic cores and electron clouds in order to regularize the singular
dipole interaction.  The Na$^+$ ions of the metal cluster interact
with the Ar dipoles predominantly by their charge. The small dipole
polarizability of the Na$^+$ core is neglected. The cluster electrons
do also couple naturally to the Coulomb field generated by the atoms.

The polarization potentials describe the long-range Cou\-lomb part of
the interactions. There remains to account for the short-range
repulsion. The repulsive Na$^+$-Ar potential is taken from
\cite{Rez95}.  The pseudo-potential for the electron-Ar core repulsion
has been modeled in the form of \cite{Dup96}, slightly readjusted by a
final fine-tuning to the NaAr molecule (bond length, binding energy,
and optical excitation). For the atom-atom interactions, we use a
standard Lennard-Jones potential.
A Van der Waals interaction is added, computed via the variance of
dipole operators \cite{Ger04b,Feh05c,Dup96}. It provides a
contribution to the long-range interaction which is crucial to produce
the faint binding of the ground states for small Na-Ar systems. Its
dynamical effects stay merely at a quantitative level.

We use standard methods \cite{Cal00} for the numerical solutions. We
solve the (time-dependent) Kohn-Sham equations for the cluster
electrons on a grid in coordinate space, using time-splitting for the
time propagation and accelerated gradient iterations for the
stationary solution. In the present calculations, we furthermore use
the cylindrically-averaged pseudo-potential scheme (CAPS) as an
approximation for the electronic wavefunctions \cite{Mon94a,Mon95a}.
Both the Na ions as well as Ar cores and valence clouds are treated by
classical molecular dynamics (MD) in full 3D. We have checked that a
full 3D treatment of the electronic wavefunctions leads to similar
results for the chosen systems which stay all close to axial symmetry.

\begin{table}[ht]
\begin{center}
\begin{tabular}{|l|l|l|l|}\hline
System&Na--Na dimer & bulk Ar--Ar & Na--Ar dimer\\
\hline
Bond energy &800 meV & 50 meV & 5 meV\\
\hline
\end{tabular}
\caption{Bond energies of dimers entering the calculations 
discussed in the text.
\label{t:dimer}}
\end{center}
\end{table}
Table \ref{t:dimer} shows the binding energies for the three possible
dimer combinations. There is a clear hierarchy of binding where the Na
cluster is strongest bound, the Ar cluster one order of magnitude less
bound, and the Na-Ar binding another order of magnitude smaller.
This leads us to expect a fragile attachment of the two clusters to each
other while the clusters as such remain intact with some readjustment
of shape, particularly at the Ar site.

\begin{figure}[htbp]
\begin{center}
\epsfig{file=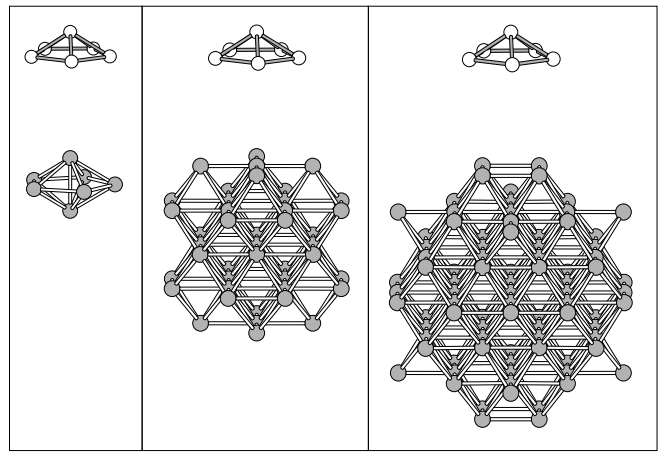,width=\linewidth}
\caption{
Initial configurations for the deposition of Na$_6$ (white balls) on
Ar$_7$ (left), Ar$_{43}$ (middle), and Ar$_{87}$ (right).
\label{fig:config}
}
\end{center}
\end{figure}

Three Ar clusters of different sizes are used as
``substrate'', Ar$_7$, Ar$_{43}$, and Ar$_{87}$. The structures of
these clusters are optimized for the given model by simulated
annealing. The collision partners are various Na clusters
taken in their ground state structure, again obtained by simulated 
annealing but this time coupled to static Kohn-Sham iterations 
for the electrons.
Most of the results presented below 
have been obtained for the strongly oblate Na$_6$
cluster, composed of a ring of five ions and an outer ion, as can be
seen from Fig.~\ref{fig:config}. Top configuration refers to the
case where the pentagon hits first the Ar cluster. Reverse
configuration corresponds to a Na$_6$ up-side down. We also consider a
few other clusters of comparable size but different shapes, especially
Na$_7$ and Na$_8$. In Na$_7$, the central pentagon is complemented by
one top and, symmetrically, one bottom ion, which altogether delivers
a much less oblate shape than Na$_6$. The Na$_8$ cluster, in turn, is
built as two layers of four ions. Mind it is electronically close to
sphericity because it contains the magic number of eight electrons.
And indeed, the shape of free Na$_8$ is, as far as it can be for a
small finite cluster, close to spherical.  In all cases, we take care
of orienting the Ar$_{N}$ relative to Na$_6$ in such a way that it
presents to the Na cluster the largest possible planar "surface". This
can be seen in Fig.~\ref{fig:config} where the initial configuration
for the collision between Na$_6$ and Ar$_7$, Ar$_{43}$ and Ar$_{87}$
are displayed~: The metal cluster initially faces respectively 1, 5,
and 4 Ar atoms. Note that in the largest system, the top Na ion starts
above an interstitial position, whereas it is initially placed above
an Ar atom in the two other cases.

The dynamics is started by giving the system a relative boost, with a "substrate" at
0~K temperature. 
We have studied the effect of "substrate" temperature 
up to 50~K. 
%
No significant differences are observed what the deposition mechanisms
and associated energy transfers are concerned, although the highest
temperature represents a thermal energy of the same order of 
magnitude (about a factor 2-3 smaller) than the 
kinetic energy in Na. 
This shows  
that relevant effects 
are primarily of potential nature, especially the balance between
short range (Ar core) repulsion and long range (Ar polarizability)
attraction, both aspects well taken into account in our model. Note
also that experimentally, deposition of metal clusters on rare gas
substrates are performed at very low temperature (typically below
25~K)~\cite{Lau03,Con06}. 
Using just 0~K temperature is thus fully sufficient for the
present aim of a qualitative study.

\section{Analysis of deposition mechanism}
\label{s:Na6}

\subsection{A visual example}
\label{s:exNa6}

\begin{figure}[htbp]
\begin{center}
\epsfig{file=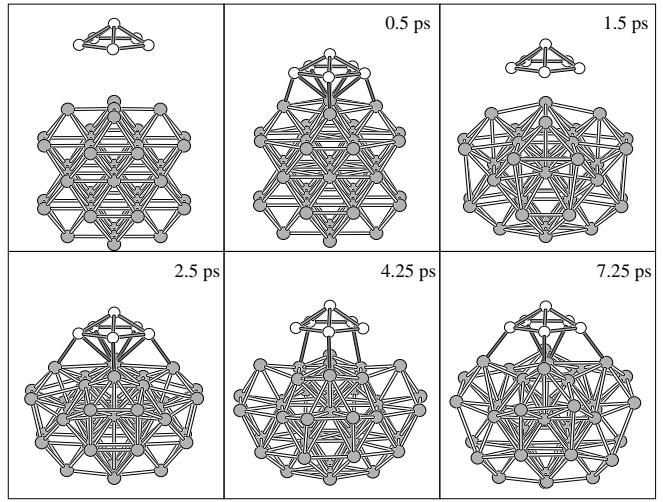,width=\linewidth}
\caption{
Snapshots of the deposition of Na$_6$ on a planar site of Ar$_{43}$,
with an initial kinetic energy of 13.6 meV/ion. Time slots are
indicated on each panel.
\label{fig:example}
}
\end{center}
\end{figure}

We first consider in Fig.~\ref{fig:example} a typical deposition
scenario for the example of Na$_6$ on Ar$_{43}$ with initial kinetic
energy $E_{\rm kin0}=13.6$ meV per Na atom.
The  Na$_6$ cluster is 
accelerated when coming closer to the Ar cluster, 
a long range polarization effect which is counterbalanced by the 
core repulsion only at very short distance. 
The pinning process proceeds stepwise with several bounces before the
metal cluster is finally attached to the rare gas cluster.  The
overall shape of the Na cluster remains basically unchanged while the
Ar cluster is somewhat rearranged, as was expected from the relative
binding energies in Table~\ref{t:dimer}.  It should be noted that the
initial kinetic energy per atom of the impinging cluster is smaller
than the bonding energies of both Na$_2$ and Ar$_2$ dimers but about
2.5 larger than the binding energy of the NaAr dimer.  Not
surprisingly neither pure Na nor pure Ar bonds are broken, but in
spite of the faintness of the NaAr bond one observes asymptotic
stitching.  This implies that an efficient kinetic energy transfer has
taken place to cool down the Na-Ar interface, the excess kinetic
energy being transferred to internal energy (potential and heat) of
the Ar and/or Na clusters. A part of the potential energy is visibly
used to reshape the Ar surface, in order to provide a more convenient contact
plane for the Na$_6$ cluster (see late times in the figure). The
energy transfer mechanisms may a priori depend both on the initial
kinetic energy and on the Ar cluster size. This will be analyzed in
the coming sections.

\subsection{Systematics on Ar cluster size}
\label{s:ArN}

\begin{figure}[htbp]
\begin{center}
\epsfig{file=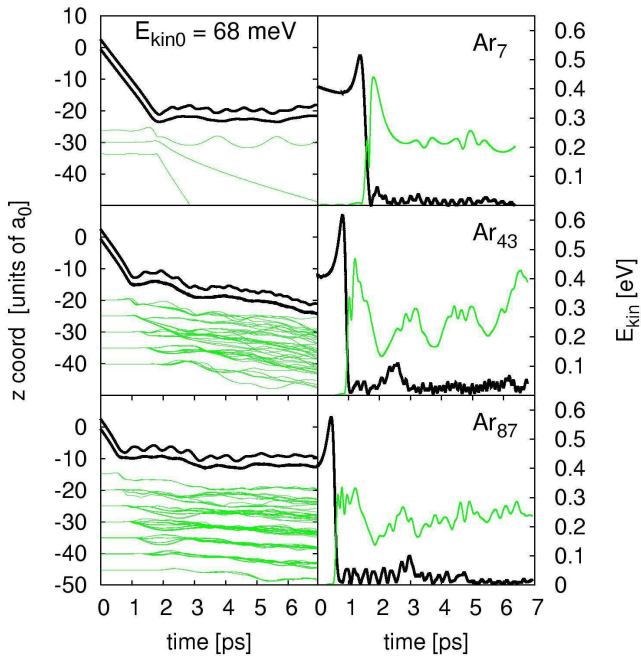,width=\linewidth}
\caption{Collision of Na$_6$ (thick lines) with an initial kinetic
  energy of 68 meV/ion, on Ar$_N$ (thin curves) for $N=7$ (top), $N=43$
  (middle) and $N=87$ (bottom). Left panels:
  $z$-coordinates; right panels: total kinetic energies, as a function
  of time.} 
\label{fig:Ek0.005Ry}
\end{center}
\end{figure}

We now turn to the question of the influence of the Ar cluster size and its
heat capacity on the deposition process.  Fig.~\ref{fig:Ek0.005Ry}
shows results for Na$_6$ deposited on various Ar clusters all for the same
initial kinetic energy of $E_{\rm kin0}=68$ meV per Na atom.  The
separation between the centers of mass of the two cluster is initially
30~$a_0$ for all systems.
The left panels show the detailed $z$ coordinates (symmetry axis and
direction of collision) for Na ions and Ar atoms. It is obvious that,
at this projectile energy, the Ar$_7$ cluster is broken into pieces
after the impact while both Ar$_{43}$ and Ar$_{87}$ are massive enough
to resist the impact. Thereby the smaller  Ar$_{43}$ is visibly more
perturbed (molten?) while the heaviest sample  Ar$_{87}$ even
maintains its shell structure. 
The kinetic energies shown in the right panels explain that trend.
Almost the same total kinetic energy of about 0.3 eV is transferred in
all three cases, independent from the Ar$_N$ size. Distributing that
equally over the Ar atoms associates a typical temperature to the Ar
cluster of about 200 K for Ar$_7$, 30 K for Ar$_{43}$, and 10 K for
Ar$_{87}$, which fits perfectly to the observation of break-up,
melting and stability. 
The largest sample thus serves as a reasonable model for deposit on a
surface in this exploratory study. However true bulk is, of
course, thermally very inert due to its large calorific capacity and one would expect
even smaller values of temperature in the case of bulk deposition under the same 
kinematic conditions.
%

A closer look to the kinetic energies in Fig.~\ref{fig:Ek0.005Ry}
shows a few interesting differences. The common feature is an initial
energy transfer which leaves about half of the initial energy, i.e.
around 0.3 eV, as kinetic energy in the Ar and very little kinetic
energy for the Na cluster. The other half is
invested in potential energy to provide the large spatial
rearrangements. Differences are seen in further evolution. For Ar$_7$
all rearrangement is finished after initial encounter and the kinetic
energies remain unchanged later on. The severe perturbation of the
medium sized Ar$_{43}$ leads to ongoing rearrangements with long
lasting exchange between potential and kinetic energy of the Ar
cluster. As a side effect, there occurs also some re-heating of
the Na$_6$ cluster. Smaller reflow of potential into kinetic energy
is seen for the heaviest Ar$_{87}$.

Fig.~\ref{fig:Ek0.005Ry} also spans very different time scales. The
energy transfer at the moment of impact is surprisingly fast. The Na
kinetic energy is fully given up within less than 0.5 ps and the
energy distribution within the Ar cluster proceeds as a {sound} wave.
This can be seen from the $z$ coordinates of the Ar atoms in the left
panels {below 2 ps}. The perturbation propagates like a straight
line through the Ar layers with a speed of 20-30 a$_0$/ps. 
Long time scales, on the other hand, remain for the final relaxation
processes. The successfully captured Na$_6$ cluster continues to
bounce and oscillate with a relaxation rate of order of 10 ps.
The fine-tuning of the Ar cluster shape seems to proceed at about
similar slow scale. 
%
Note, however, that this shape evolution initiated by the sound wave
might differ for an infinite substrate where the
sound wave would dissolve into deep layers.
In the small finite systems here, the wave is dissipated by diffuse scattering from the
opposite surfaces of the Ar cluster which turns collective energy into
heat. A large part of the Ar kinetic energy seen in
Fig.~\ref{fig:Ek0.005Ry} thus becomes thermal energy. 
The case of 
largest Ar cluster, Ar$_{87}$, behave somewhat different.
It presents cleaner surfaces as compared
to smaller clusters and it shows a faint reverberation of
the wave when the latter reaches the edge of the cluster, as is
observed in 
the lower left panel of Fig.~\ref{fig:Ek0.005Ry}. The reflected
wave bounces back
with almost the same velocity to the side which qas hit initially.
There it
returns some momentum to the Na$_6$, as can be seen in the small revival of
its kinetic energy at about 3 ps (lower right panel). This reverbation is, of course,
due to the finite size of Ar$_{87}$ and would not appear in the case
of an infinite substrate. Nevertheless, from the metal cluster point of
view, this reflection seems to change very little its final shape and
distance to the Ar ``surface''. In that sense, this system still
mimics
the gross features of
deposition on an infinite surface.

\subsection{Systematics on initial kinetic energy}
\label{s:Ekin0}

\begin{figure}[htbp]
\begin{center}
\epsfig{file=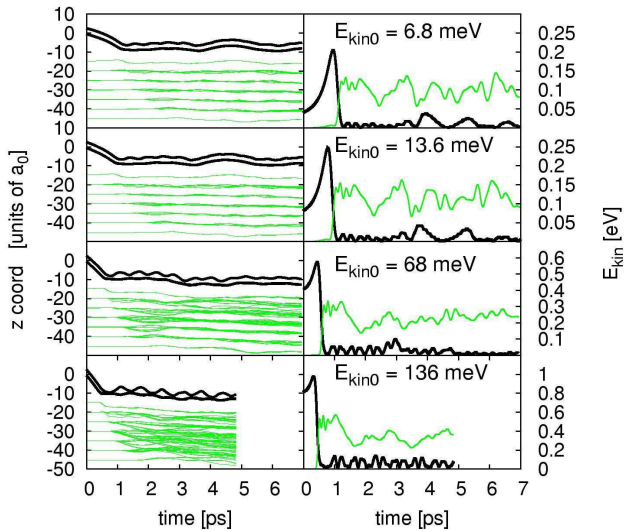,width=\linewidth}
\caption{$z$-coordinates (left) and total kinetic energies (right), as
  a function of time, in the collision of Na$_6$ (thick curves) on
  Ar$_{87}$ (thin curves),
  for four different initial kinetic energies $E_{\rm kin0}$.}
\label{fig:na6ar87}
\end{center}
\end{figure}

It is expected that the deposition depends on the velo\-city of the
projectile. We analyze this effect in the case of Na$_6$ deposition on
Ar$_{87}$, with varying $E_{\rm kin0}$. Results are plotted in
Fig.~\ref{fig:na6ar87} for $E_{\rm kin0}$ between 6.8 meV and 136 meV
per Na atom.  The striking feature is the overall similarity of all
the different cases. This concerns in particular the very fast and
almost complete energy transfer from the Na kinetic energy to the
other degrees of freedom. A common feature is also the quick capture
of the Na cluster and its long standing residual oscillations. 
Of course details of the scenarios differ from one case to the next.
The perturbation  of the Ar cluster increases naturally with increasing
initial energy. There is also an interesting effect at initial times.
The kinetic energy of the Na cluster first increases before contact.
That is due to the medium range polarization interaction which is
attractive. The additional acceleration depends on the initial
velocity. More energy is gained for the slower velocities because the
cluster moves for a longer time in the attractive  regime. Thus the
kinetic energy is almost quadrupled for the lowest $E_{\rm kin0}=6.8$ meV
while only 10\% effect is left over for the fastest collision here.
Besides these subtle differences in detail, the energy deposit
proceeds in all cases the same way, namely in an extremely quick
fashion leaving basically no residual kinetic energy at the side of
the Na cluster projectile. One thus finds that in a relatively large
range of projectile kinetic energies, the Na projectile is glued to
the Ar target.  Except again for oscillations in the most energetic
case, the overall structure and position (with respect to the Ar
target) of the deposited Na$_6$ seems to depend very little on the
initial energy.

\section{Time evolution of shapes}
\label{s:wet}

\subsection{Cluster distance}
\label{s:shape}

The above calculations show that the relaxation times of the whole
deposition process goes beyond times we can reasonably
simulate. However, we have seen from Figs.~\ref{fig:Ek0.005Ry} and
\ref{fig:na6ar87} that around 6--7 ps the kinetic energy of the Ar
cluster and the cluster distance seems to reach a constant mean value.
The closeness of the two clusters can be quantified 
in terms of a ``deposit coordinate'' $d$ that we define as
\begin{equation}
\label{d}
 d = \left[ \frac{1} {N_{\rm Na} N_{\rm Ar}} \sum_{I,a}
 \frac{1}{ | \mathbf R_I - \mathbf R_a|^4 } \right]^{-1/4} ,
\end{equation}
where the index $I$ (respectively $a$) refers to the Na ion
(respectively Ar atom) cores and $\mathbf R_I$ (respectively $\mathbf
R_a$) to their position. The inverse puts weight on the closest
partners. The actual power 4 in Eq.(\ref{d}) is a matter of decision:
we calculated with powers $n=1$ up to $n=7$ and found that $n=4$
provides a good compromise between averaging and details.

\begin{figure}[htbp]
\begin{center}
\epsfig{file=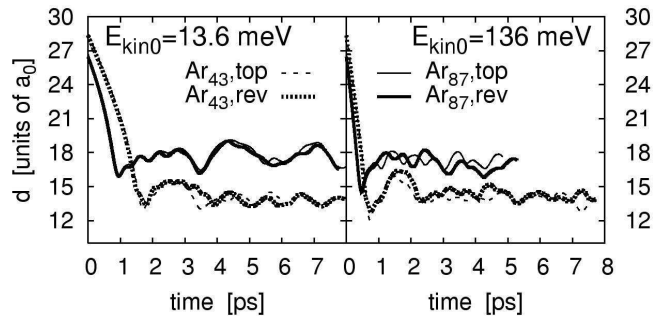,width=\linewidth}
\caption{Deposit coordinate $d$, defined in Eq.(\ref{d}), as a
  function of time, for Na$_6$@Ar$_{43}$ (dots and dashes) and
  Na$_6$@Ar$_{87}$ (full lines), with $E_{\rm kin0}=13.6$ meV/ion (left
  panel) and 136 meV/ion (right panel). Results in both systems are
  presented for the top (dashes and thin curve) and the reverse (dots
  and thick curve) configuration of the Na$_6$.}
\label{fig:d}
\end{center}
\end{figure}

Fig.~\ref{fig:d} shows the coordinate $d$ for two different sizes
of the Ar clusters and for two different $E_{\rm kin0}$, i.e. initial
Na kinetic energies per ion. The pattern are surprisingly similar for
both energies and also similar for both sizes. The Na cluster is
catched at first impact. There remain some oscillations in distance of
about 2--3 a$_0$ which relax very slowly (order 10 ps).  There is a
global difference to the extend that the distance is closer for
Na$_6$@Ar$_{43}$. This is due to the larger rearrangements in that
case which, in turn, provide a better contact area for the Na$_6$.

We have also varied the orientation of the Na cluster. The standard
configuration for Na$_6$ was such that the ring of five Na ions stood
closer to the surface while the top ion was pointing away (the
situation denoted as ``top'' in the figure). Within axial symmetry,
there is also the reverse situation where the top ion is facing
towards the surface (denoted ``rev'' in the figure).  The result for
reverse initial configurations are also shown in Fig.~\ref{fig:d}.
They are almost indistinguishable from the top configuration. This
indicates that the Na$_6$ remains basically unchanged in all cases.
This will be corroborated in the next section studying shape dynamics.

\subsection{Global moments}
\label{ss:shape}

The results discussed up to now indicate that there may be substantial
rearrangements at the side of the Ar substrate.  Less is yet known
about the shape of the Na cluster: do we have a soft landing, some
plastic deformation or a wetting behavior? In order to quantify these
questions we have performed a shape analysis in terms of the first three
multipole moments both for Na$_6$ and Ar$_N$. These moments are given
by
\begin{eqnarray*}
\sqrt{\langle r^2 \rangle} &=& \sqrt{\langle x^2 + y^2 + z^2 \rangle} \cr
                           &=&  \left( \frac{1}{p} \sum_{i=1}^p ({x_i}^2
     + {y_i}^2 + {z_i}^2) \right)^{1/2},\cr 
\beta_2 &=& \sqrt{\frac{\pi}{5}} \frac{1}{\langle r^2 \rangle}
\langle 2 z^2 - x^2 - y^2\rangle, \cr
\beta_3 &=& \left(\frac{2}{5 \langle r^2 \rangle} \right)^{3/2}
\langle z\left(z^2 - \frac{3}{2}(x^2+y^2) \right)\rangle,
\end{eqnarray*}
where $p$ is either the number of Na atoms $N_{\rm Na}$ or the number
of Ar atoms $N_{\rm Ar}$, and $x$, $y$ and $z$ are the coordinates of
the Na (Ar) 
atom with respect to the center of mass of the Na (Ar) cluster.
The r.m.s. radius $r$ stands for the overall extension (monopole
moment) and the deformations are parameterized as dimensionless
quantities which have immediate geometrical  meaning independent of
system size. For example, a value of $|\beta_2|\approx 0.8$ is a large
quadrupole deformation with axis ratio of about 2:1.

\begin{figure}[htbp]
\begin{center}
\epsfig{file=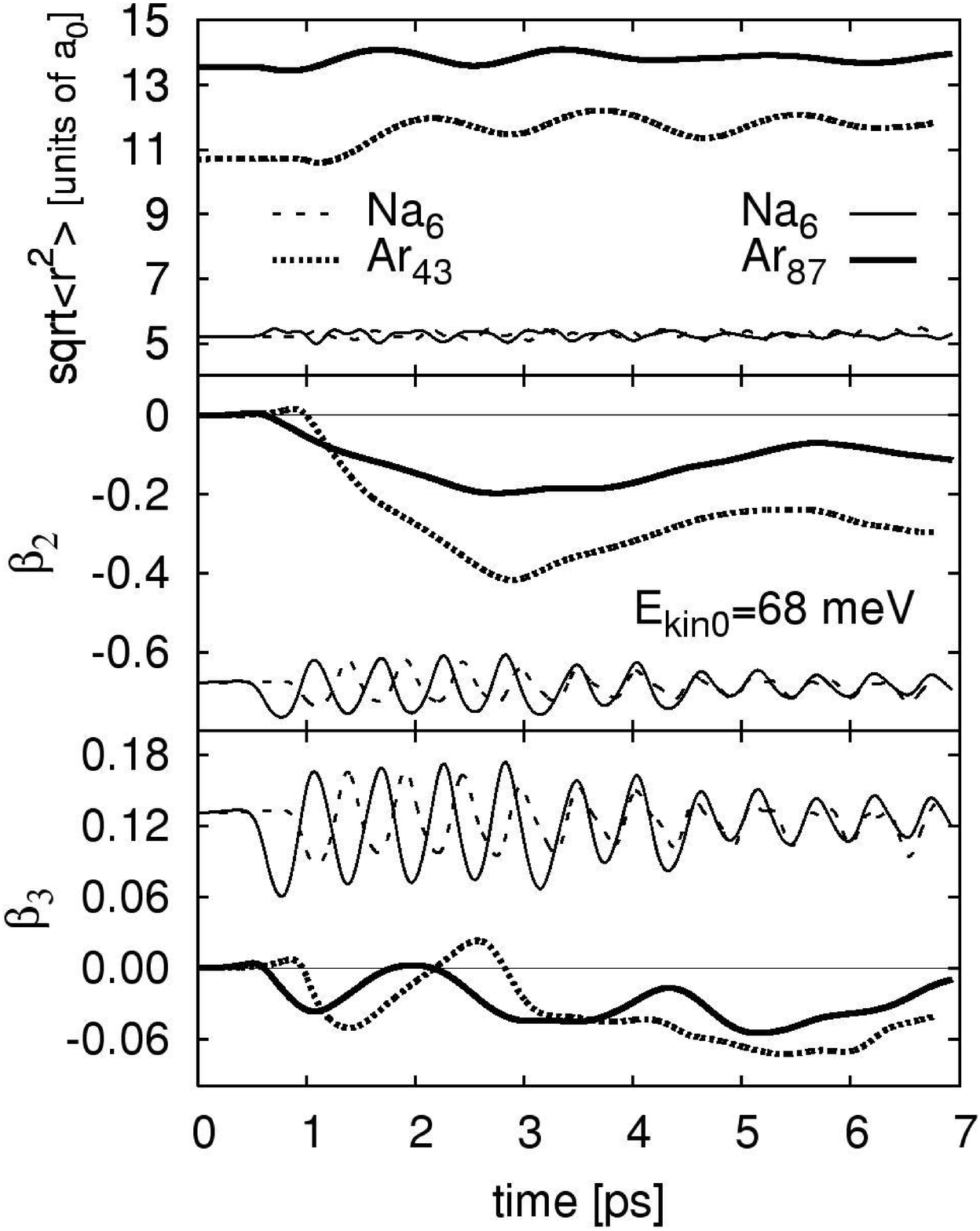,width=0.8\linewidth}
\caption{Three first multipole moments $\sqrt{\langle r^2 \rangle}$,
  $\beta_2$ and $\beta_3$, as a function of time, for Na$_6$ (thin
  dashes and lines) deposited on Ar$_{43}$ (thick dots) and Ar$_{87}$
  (thick full lines) for $E_{\rm kin0}=68$ meV/ion.}
\label{fig:size_moments}
\end{center}
\end{figure}

Fig.~\ref{fig:size_moments} shows the three momenta for the Na and
Ar subsystem in the cases Na$_6$@Ar$_{43}$ and Na$_6$@Ar$_{87}$ for
the moderate initial kinetic energy $E_{\rm kin0}=68$ meV per Na ion.
The shape of Na$_6$ is rather rigid in any case. 
There are some deformation oscillations short after impact which 
relax within about 3 ps. These oscillations are predominantly caused
by the outer ion. The ring is tightly bound and stays more robust.
Note that the relaxation is much faster than for the overall 
bouncing oscillations of the cluster (see Fig.~\ref{fig:d}).
That is related to the binding properties shown in
Table~\ref{t:dimer}, that is, the NaAr binding is softer.
The Ar clusters, after the impact with Na$_6$, increase slightly in
size due to the heating by energy absorption. The growth is relatively
larger for the smaller Ar$_{43}$ which acquires a higher temperature
as discussed above.
The Ar clusters do also undergo a strong persistent change in
deformation towards a sizeable oblate and somewhat pear-like shape. They
obviously accommodate their configuration as to establish a most
compact combined system.

\begin{figure}[htbp]
\begin{center}
\epsfig{file=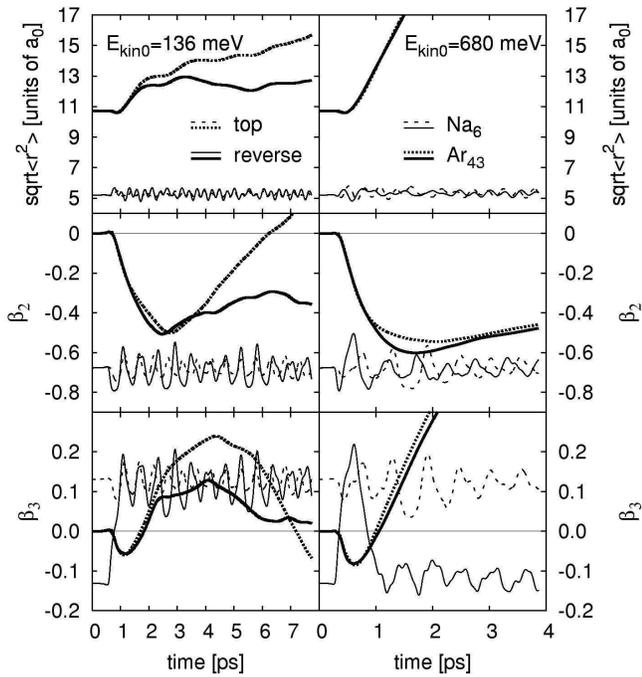,width=\linewidth}
\caption{Three first multipole moments, as a function of time, for
  Na$_6$ (thin curves) 
  deposited on Ar$_{43}$ (thick curves), with $E_{\rm kin0}=136$ meV/ion
  (left) and 680 meV/ion (right). On each panel are compared the moments
  obtained from the top (dots and dashes) and reverse (full lines)
  configurations.}
\label{fig:Ekin_moments}
\end{center}
\end{figure}

For the case of Na$_6$@Ar$_{43}$, the effect of initial kinetic energy
and orientation (Na cluster in ``top'' or ``rev'' configuration, see
section~\ref{s:shape}) are shown in Fig.~\ref{fig:Ekin_moments}. For
the large $E_{\rm kin0}=680$ meV/ion (right 
panels), the Ar$_{43}$ emits rapidly several atoms, thus yielding a
rapidly increasing radius and octupole moment while a more moderate
evolution emerges for the less violent $E_{\rm kin0}=136$ meV/ion.  The
collision with the Na$_6$ top configuration is more violent because
the whole ring bounces first and at once. This gives rise to an even
higher Ar deformation for both initial energies.
However, at the side of Na$_6$, mean value and amplitude of the shape
oscillations are similar in all cases, for both configurations and for
both $E_{\rm kin0}$. Note the change of sign of $\beta_3$ for the
reverse configuration. This indicates that the outer ion oscillates
through the pentagon, within the given time once for $E_{\rm
kin0}=136$ meV and twice at 680 meV. But the other two moments stay
very robust in spite of the violence unload upon the Ar$_{43}$
subsystem.

\subsection{Wetting behavior; comparison with Na$_{7}$ and Na$_8$}
\label{s:Na78}

We thus find that the Na$_6$ is very robust under any conditions when
deposited on the rather soft Ar material.  It maintains its oblate
shape and overall radius.  The top ion is more loosely bound and may
undergo larger oscillations which become apparent in the octupole
momentum. The results for Na$_6$ thus suggest that wetting of the
surface is rather unlikely in that combination of materials. However,
Na$_6$ is not a very conclusive test case as it has already a close to
planar structure. One ion still sticking out of the surface plane
looks not so dramatic. In order to countercheck, we considered two
neighboring clusters, namely Na$_7$ (less oblate than Na$_6$) and the
almost spherical Na$_8$.
\begin{figure}[htbp]
\begin{center}
\epsfig{file=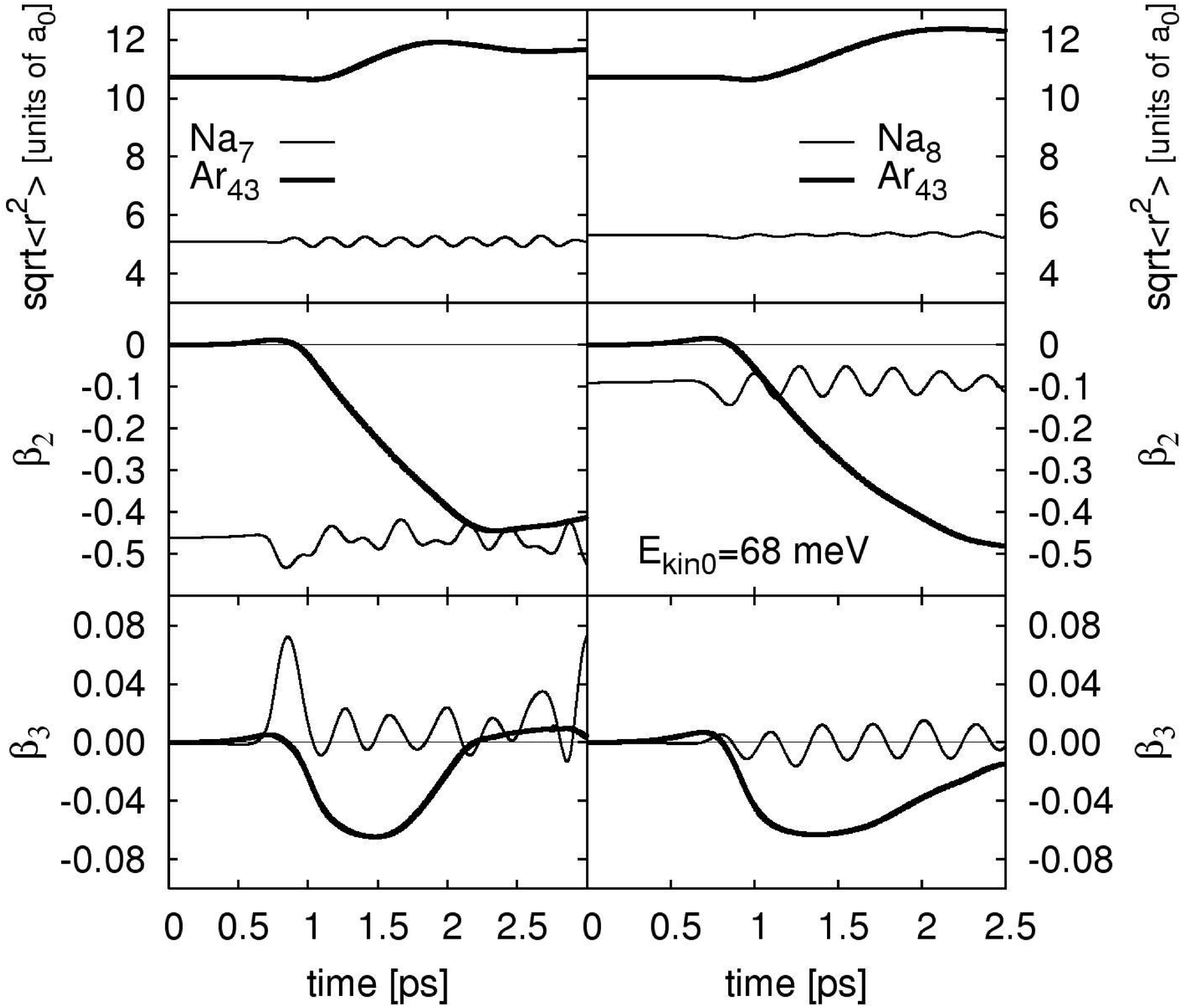,width=\linewidth}
\caption{The three multipole moments, as a function of time, for
  Na$_7$ (thin curves, left panels) and Na$_8$ (thin curves, right panel)
  deposited on Ar$_{43}$ (thick curves), with $E_{\rm kin0}=68$ meV/ion. 
}
\label{fig:Ekin_moments_78}
\end{center}
\end{figure}
The results for the multipole moments are shown in
Fig.~\ref{fig:Ekin_moments_78}. They are very similar to what we have
seen 
already for collisions with Na$_6$. The Na$_7$ or Na$_8$ clusters
maintain basically their shape while the Ar basis undergoes a
substantial rearrangement to create space for landing. There is not
the slightest hint of a wetting. 
There are, of course, small shape oscillations in the Na clusters due
to the internal excitation at the time of impact. It is interesting to
note that these oscillations differ in detail between the two
clusters. The Na$_8$ with its magic electron number 8 seems to be more
rigid and shows generally less oscillation amplitude.

\section{Conclusion}
\label{s:ccl}

In this paper we have investigated slow collisions of a small Na
cluster with a large Ar cluster as a test case for deposit on an Ar
surface. To that end, we have employed a hierarchical model treating
the Na cluster in full detail by time-dependent density-functional
theory and the Ar atoms at a thoroughly classical level with position
and polarization as dynamical variables. We have studied in detail
the influence of various parameters on the deposition scenario, namely
initial collision energy, Ar cluster size, and relative orientation of
the Na cluster.

We have found that, except for the case of the smallest Ar system, the
basic scenario is robust in the sense that it depends very little on
the collision parameters. The Na clusters are well bound while Ar
binding is much softer and the Na-Ar binding even less. As a
consequence, the Ar substrate constitutes a very efficient shock
absorber. The impinging Na cluster is stopped immediately at first
impact and attached to the Ar system while the latter takes over
practically all initial Na energy. It seems impossible to tune
conditions under which the Na cluster is reflected. Enhancing the
initial energy rather leads to destruction of the Ar system.
The initial large energy transfer from the Na cluster to the Ar system
is very quick, taking less than 0.5 ps.  The transferred energy is
distributed also very quickly over all Ar atoms, propagating like a
shock wave with speed of sound through the medium. About half of the
energy goes into potential energy and part of it is used up for a 
large spatial rearrangement of the Ar cluster which aims to provide a
most compact combined system. These rearrangements proceed on a slower
time scale of 4--10 ps. 
At the side of the Na cluster, we have observed that in all cases its
shape is little affected by the dynamical processes during deposit.
It always safely attaches to the surface but only loosely bound and
keeps nearly its free shape. The interface interaction is too weak to
induce a wetting mechanism for this material combination.

Calculations with larger clusters and planar surfaces for the same
combination confirm the above findings. It is now most interesting to
check other combinations of materials. Work in that direction is in
progress.

Acknowledgments: 
This work was supported by the DFG, project nr. RE 322/10-1,
the french-german exchange program
PROCOPE nr.  04670PG, the CNRS Programme "Mat\'eriaux" 
(CPR-ISMIR), Institut Universitaire de France, and the
Humbodlt foundation,
and has benefit from the
Calmip (CAlcul en MIdi-Pyrénées) computational facilities.

\end{document}